\documentclass{Lisbon}

\newcommand{\vpint}{\int\makebox[0mm][r]{\bf --\hspace*{0.13cm}}}
\newcommand{\ds}{\displaystyle}
\newcommand{\vph}{\varphi}
\title{\bf Strong hadronic decays in $QCD_2$.}
\author{Yu.S.Kalashnikova and A.V.Nefediev}
\begin{document}
\maketitle

\abstracts{Strong hadronic decays are discussed in the 't~Hooft model for the
two-dimensional QCD in the limit of infinite number of colours. Adler
selfconsistency condition is derived for decays with pions in 
the final state and the corresponding amplitude is demonstrated to
vanish identically in the chiral limit.}

The 't~Hooft model for $QCD$ in two dimensions in the limit of infinite
number of colours \cite{'tHooft} is known to be a
good laboratory for testing many phenomena which take place in $QCD_4$, among those are confinement and the chiral 
symmetry breaking. The Lagrangian of $QCD_2$ has the form
\be
L(x)= -\frac{1}{4}F^a_{\mu\nu}(x)F^a_{\mu\nu}(x)+\bar q(x)(i\hat{D}-m)q(x),
\ee
and we fix the axial gauge $A_1^a=0$ for the gluonic field.
Besides that we use the principal value prescription 
for regularizing the infrared behaviour of the gluonic propagator
\cite{Bars&Green} and our convention for
$\gamma$-matrices is $\gamma_0 = \sigma_3$,
$\gamma_1 = i\sigma_2$, $\gamma_5 = \gamma_0\gamma_1$. The large $N_C$
limit implies that $\gamma=\frac{g^2N_C}{4\pi}$ remains finite and plays the
role of the effective coupling constant. 

The Hamiltonian of the model in the given gauge 
can be obtained by the standard methods and reads (see \cite{attacks} for similar approach to
$QCD_4$)
\be
H=\int dx q^{+}(x)\left(-i\gamma_5\frac{\partial}{\partial x}+m\gamma_0\right)
q(x)-\frac{g^2}{2}\int dxdy\rho^a(x)\frac{\left|x-y\right|}{2}\rho^a(y),
\label{H}
\ee
where $\rho^a(x)=q^{+}(x)t^aq(x)$ and the quark dressed field is
\be
q_{\alpha}(x_0,x)=\int\frac{dk}{2\pi}\left[u_i(k)b_{\alpha}(x_0,k)+v_i(-k)d_{\alpha}(x_0,-k)\right]e^{ikx}.
\label{qi}
\ee
The fermionic vacuum is defined as $b_{\alpha}(k)|0\rangle= d_{\alpha}(-k)|0\rangle =0$
whereas
the amplitudes $u$ and $v$ are given by
\be
u(k)=T(k)\left(1 \atop 0 \right)\quad v(-k)=T(k)\left(0 \atop 1 \right)\quad
T(k)=e^{-\frac{1}{2}\theta(k)\gamma_1}.
\ee

Parameter $\theta(k)$ has the meaning of the Bogoliubov--Valatin angle
describing rotation from bare to dressed quarks.
Following \cite{Bars&Green} Hamiltonian (\ref{H}) can be normally
ordered in the basis of fermionic operators $b$ and $d$:
\be
H=LN_C{\cal E}_v+:H_2:+:H_4:,
\ee
where ${\cal E}_v$ is the vacuum energy density ($L$ being the length
of the space), $:H_2:$ is bilinear in quark operators and $:H_4:$ contains
the fourth powers.

The standard requirement that $:H_2:$ has the diagonal form
\be
:H_2:=\int\frac{dk}{2\pi}
E(k)\left\{b_{\alpha}^{+}(k)b_{\alpha}(k)+d_{\alpha}^{+}(-k)d_{\alpha}(-k)\right\},
\ee
where $E(k)$ is the dressed quark dispersive law, leads to the gap equation 
\cite{Bars&Green}
\be
p\cos\theta(p)-m\sin\theta(p)=\frac{\gamma}{2}\vpint\frac{dk}{(p-k)^2}\sin[\theta(p)-\theta(k)],
\label{gap}
\ee
which was studied numerically in \cite{Ming Li} and a nontrivial
solution surviving in the chiral limit, was found. 

As the next step we perform diagonalization of the model in the sector of bound
states --- $q\bar q$ mesons.
To this end we take $:H_4:$ part of the Hamiltonian into account
and apply another Bogoliubov transformation generalized for the case of 
compound operators \cite{KNV}. Then we find the following operator annihilating
mesonic state
\be
m_n=\int\frac{dq}{2\pi\sqrt{N_C}}\left\{d_i(P-q)b_i(q)\vph_+^n(q,P)+
b^{+}_i(q)d^{+}_i(P-q)\vph_-^n(q,P)\right\},
\label{m}
\ee
and a similar expression for $m^+$, which possess
the standard bosonic commutation relations 
and diagonalize the Hamiltonian of the theory
\be
H=LN_C{\cal E}'_v
+\sum\limits_{n=0}^{+\infty}\int\frac{dP}{2\pi}P^0_n(P)m^+_n(P)m_n(P)+
O\left(\frac{1}{\sqrt{N_C}}\right).
\label{HHHH}
\ee

The two wave-functions $\varphi_{\pm}$ describing forward and backward in time motion
of the $q\bar q$ pair inside the meson are solutions to the bound-state equation~\cite{Bars&Green}
\be
\left\{
\begin{array}{c}
[K(p,P)-P_0]\vph_+(p,P)=\gamma\ds\vpint\frac{\ds dk}{\ds (p-k)^2}
\left[C\vph_+(k,P)-S\vph_-(k,P)\right]\\

[K(p,P)+P_0]\vph_-(p,P)=\gamma\ds\vpint\frac{\ds dk}{\ds (p-k)^2}
\left[C\vph_-(k,P)-S\vph_+(k,P)\right],
\end{array}
\right.
\label{BG}
\ee
where $K(p,P)=E(p)+E(P-p)$, $C=\cos\frac{\theta(p)-\theta(k)}{2}
\cos\frac{\theta(P-p)-\theta(P-k)}{2}$, $S=\sin\frac{\theta(p)-\theta(k)}{2}
\sin\frac{\theta(P-p)-\theta(P-k)}{2}$. They also obey the Bogoliubov-type orthonormality
conditions
\be
\begin{array}{rcl}
\ds\int\frac{\ds dp}{\ds 2\pi}\left(\vph_+^n(p,P)\vph_+^{m}(p,P)-\vph_-^n(p,P)\vph_-^m(p,P)
\right)&=&\delta_{nm}\\
\ds\int\frac{dp}{2\pi}\left(\vph_+^n(p,P)\vph_-^{m}(p,P)-\vph_-^n(p,P)\vph_+^m(p,P)
\right)&=&0.
\label{norms}
\end{array}
\ee

One of the most interesting features of the bound-state equation (\ref{BG}) is
its massless solution present in the spectrum in the chiral limit. It reads \cite{KNV}
\be
\varphi^{\pi}_{\pm}(p,P) = \sqrt{\frac{\pi}{2P}}\left(\cos\frac{\theta(P-p)-\theta(p)}{2}
\pm \sin\frac{\theta(P-p)+\theta(p)}{2}\right).
\label{11}
\ee

Note that $\varphi^{\pi}_-(p,P\to 0)\sim\varphi^{\pi}_+(p,P\to 0)$, so that the $q\bar q$
pair in the pion spends half time in the backward in time motion, thus both
components of the pion wave-function are equally important and should be taken
into account.

To proceed further we introduce the matrix wave-function \cite{Bars&Green}
\be
\Phi(p,P) = T(p)\left(\frac{1+\gamma_0}{2}\gamma_5\vph_+(p,P) +
\frac{1-\gamma_0}{2}\gamma_5\vph_-(p,P)\right)T^+(P-p),
\label{phi}
\ee
and the incoming and outgoing matrix meson-quark-antiquark vertices \cite{KN}
\be
\Gamma_M(p,P)=\int\frac{dk}{2\pi}\gamma_0\frac{\Phi_M(k,P)}{(p-k)^2}\gamma_0\quad
\bar \Gamma_M(p,P)=\gamma_0\Gamma_M^+(p,P)\gamma_0.
\label{14}
\ee

Taking advantage of the fact that the full quark-antiquark scattering 
amplitude is known explicitly \cite{KN}, one can derive the 
following Ward identities for the
dressed vector and axial-vector current-quark-antiquark vertices $v_{\mu}$
and $a_{\mu}$ (see \cite{Einhorn} for similar expression in the light-cone gauge):
\be
-iP_{\mu}v_{\mu}(p,P)=S^{-1}(p)-S^{-1}(p-P),
\label{25}
\ee
\be
-iP_{\mu}a_{\mu}(p,P) = S^{-1}(p)\gamma_5+\gamma_5S^{-1}(p-P),
\label{27}
\ee
where $S(p)$ is the dressed quark Green's function.

Using the explicit form of the pion vertex
which can be easily found with the help of equations (\ref{11}) -- (\ref{14}), 
one arrives to the conclusion that the pion-quark-antiquark vertex is nothing but a linear 
combination of the vector and the axial-vector currents divergences (\ref{25})
and (\ref{27}) 
\be
\Gamma_{\pi}(p,P)=S^{-1}(p)(1+\gamma_5)-(1-\gamma_5)S^{-1}(p-P)=-iP_{\mu}v_{\mu}-iP_{\mu}a_{\mu}.
\label{16}
\ee

We are now in the position to calculate the strong decay amplitude for the 
process $A\to B+C$. 
The Hamiltonian approach gives a natural environment for this task
as the Hamiltonian (\ref{HHHH}) with reconstructed sub-leading terms $O(1/\sqrt{N_C})$
immediately
reproduces the amplitude of the process. In the meantime, there is another a more elegant
way exploiting the effective diagrammatic technique. The matrix
element for the discussed decay reads
$$
M(A \ra B+C)=-\frac{i\gamma^3}{\sqrt{N_C}}\int \frac{d^2k}{(2\pi)^2}\times\hspace*{4cm}
$$
$$
\times Sp(\Gamma_A(k+P_B,P_A)S(k-P_C)\bar \Gamma_C(k,P_C)
S(k)\bar \Gamma_B(k+P_B,P_B)S(k+P_B))
$$
\be
\hspace*{7cm}+(B \leftrightarrow C).
\label{13}
\ee

One of the most interesting issues of
the hadronic spectroscopy is the role played by the pions, as Goldstone
bosons, in the hadronic decays. Inserting the pion vertex (\ref{16}) into
(\ref{13}) one immediately arrives at the conclusion that the decay amplitude
vanishes in the chiral limit if at least one of the final particles is the pion,
and this happens for any pion momentum!\footnote{The latter observation seems natural
in view of the fact that the pionic decay constant is dimensionless in D=2 thus defining no
low-energy scale for the theory\cite{KN}.} \cite{KN}
\be
M(A \ra \pi+C)\equiv 0.
\label{ac}
\ee

This relation plays the role of the Adler selfconsistency
condition in $QCD_2$~\cite{Adler}. It is entirely due
to destructive interference between the two components of the pionic wave-function
$\varphi_+^{\pi}$ and $\varphi_-^{\pi}$. The same conclusion follows immediately from
(\ref{16}) as both, vector and axial-vector currents are conserved in the 't~Hooft model in
the chiral limit. 

Financial support of RFFI grants 00-02-17836 and 00-15-96768 and INTAS-RFFI grant IR-97-232
is gratefully acknowledged.

\end{document}